\documentclass[prl,superscriptaddress,showpacs,twocolumn]{revtex4}
\usepackage{amssymb}
\usepackage{graphicx}
\usepackage{dcolumn}
\usepackage{bm}
\usepackage{amsmath}
\usepackage{tikz}

\begin{document}

\title{Large Chern Number Topological Superfluids in Coupled Layer System}
\author{Beibing Huang}
\affiliation{Department of Physics and Center for Quantum Coherence, The Chinese University of Hong Kong, Shatin, N.T., Hong Kong, China}
\affiliation{Department of Physics, Yancheng Institute of Technology, Yancheng, 224051, P. R. China} 
\author{Chun Fai Chan}
\author{Ming Gong}
\thanks{skylark.gong@gmail.com}
\affiliation{Department of Physics and Center for Quantum Coherence, The Chinese University of Hong Kong, Shatin, N.T., Hong Kong, China}

\begin{abstract}
We investigate topological superfluids in a coupled layer system, in which transitions between different topological superfluids can be realized by controlling the binding energy, interlayer tunneling and layer asymmetry {\it etc}. These topological transitions are characterized by energy gap closing and reopening at the critical points at zero momentum where the Chern number and sign of Pfaffian undergo a discontinuous change. In a hard wall boundary the bulk-edge correspondence ensures that the number of edge modes exactly equals the Chern number, whereas all the edge modes localized at the same edge propagate along the same direction. However, in a trapped potential these edge modes are spatially localized at the interfaces between different topological superfluids, where the number of edge modes exactly equals the Chern number difference between the left and right interface. These topological phases can be detected by spin texture at or near zero momentum, which changes discretely across the phase transition points due to band inversion. The model can be easily generalized to multilayer system in which the Chern number can equal any positive integer. These large Chern number topological superfluids provide fertile grounds for exploring new quantum matters in context of ultracold atoms.
\end{abstract}

\pacs{67.85.Lm, 74.20.Rp, 74.20.Fg, 71.10.Fd}
\maketitle

The experimental realization of spin-orbit coupling (SOC) in ultracold atoms\cite{SOC1, SOC2, SOC3, SOC4, SOC5, SOC6, SOC7, SOC8}, see also a recent review article \cite{REV},  opens a totally new arena for exploring exotic topological Bardeen-Cooper-Schrieffer (BCS) and Fulde-Ferrell-Larkin-Ovchinnikov (FFLO)\cite{FF64, LO64} superfluids in the context of ultracold atoms\cite{MG11, MG112, MG113, MG12, JP, KS12, LH, MI, MI13, CQU13, WZ13, XJL13, CC13, RW12, Gang, HuHuia, HuHuib, XY, HLY, HLY2}. The underlying idea is that when the chemical potential just fills one band, the effective pairing at this band should be $p$-wave type due to the momentum dependent spin polarization induced by SOC and Zeeman field. Notice that the $p$-wave pairing is a basic prototype for topological superconductors\cite{Pwave1, Pwave2, Pwave3, Pwave4}, thus topological superfluids can be expected in this new platform. The topological phase can be realized when  $\Gamma^2 > \mu^2 + \Delta^2$, where $\Gamma$, $\mu$ and $\Delta$ are the corresponding Zeeman splitting, chemical potential and order parameter respectively. Since the pairing takes place at the lower band, it can survive even when the Zeeman field is much larger than the order parameter strength --- a typical example to go beyond the Clogston-Chandrasekhar limit\cite{CC1, CC2} in the $s$-wave pairing system. This system belongs to topological class D and is characterized by topological invariant $\mathbb{Z}$\cite{ludwig, zhao}; however the Chern number in this model at most equals one, thus only one topological phase with Chern number $\mathcal{C} = 1$ can be realized. The counterpart condition is totally different in condensed matter physics such as integer quantum Hall state\cite{IQHE}, quantum anomalous Hall model\cite{Fang}, Haldane model\cite{Doru}, stacked graphene\cite{FanZhang} and $^3$He film\cite{Volvik}, where Chern number greater than one can be easily realized.

In this Letter we show that large Chern number topological superfluids can be realized in a coupled layer 
system. We utilize a bilayer system to illustrate the major idea via self-consistent calculation. In this 
model the phase transitions between different topological superfluids can be realized by controlling the binding energy, interlayer 
tunneling and layer asymmetry {\it etc}. These topological transitions are characterized by
energy gap closing and reopening at the critical points at zero momentum where the Chern number and sign of Pfaffian
undergo a discontinuous change. In a hard wall boundary the bulk-edge correspondence ensures that the number of edge modes 
exactly equals the Chern number, whereas all the edge modes localized at the same edge propagate along the same direction.
 However, in a trapped potential these edge modes are spatially localized at the interfaces between different topological superfluids, where
the number of edge modes exactly equals the Chern number difference between the left and right interface. These topological phases can be detected by 
spin texture at or near zero momentum, which also changes discretely across the critical points due to band inversion. The model can be easily 
generalized to multilayer system in which the Chern number can equal any positive integer. These large Chern number topological superfluids provide 
fertile grounds for exploring new quantum matters in context of ultracold atoms.

{\it Theoretical model}. We first consider a spin-orbit coupled bilayer system, which can be described by\cite{MG11, MG112, MG113, MG12, JP, KS12, LH, MI, MI13, CQU13, WZ13, XJL13, CC13, RW12, Gang, HuHuia, HuHuib, XY, HLY, HLY2}, $\mathcal{H} =\sum_{ {\bf k}, i=1,2, ss'} c_{{\bf k}is}^\dagger (\epsilon_{\bf k}^i + \alpha ({\bf k}_i\times \boldsymbol{\sigma})_{z} - \Gamma \sigma_z)_{ss'} c_{{\bf k}is'} - t \sum_{{\bf k}s}(c_{{\bf k}1s}^\dagger c_{{\bf k}2s} + c_{{\bf k}2s}^\dagger c_{{\bf k}1s}) + \mathcal{V}_{\text{int}}$,  where we have assumed that the two layers ($i = 1, 2$) have the same SOC strength $\alpha$ and $\epsilon_{\bf k}^i = {{\bf k}^2 \over 2m} - \mu_i$. Hereafter we let $\mu_1 = \mu$ and $\mu_2 = \mu + \delta \mu$, where $\delta \mu$ controls the asymmetry of the bilayer system. $c_{{\bf k}is}^\dagger$ is the creation operator for
the fermion particle with momentum ${\bf k} = (k_x, k_y)$ and spin $s = \uparrow, \downarrow$ and $t$ is the spin independent 
tunneling between the two layers. The tunneling and asymmetry --- two major new parameters in our model --- can be controlled 
by the laser fields along the perpendicular ($z$) direction in experiments. The last term represents the many-body 
interaction and can be formulated as $g \delta({\bf r} - {\bf r}')$. Notice
that the overlap between the wavefunctions of the two layers are suppressed
by the barrier between them, we can safely neglect the interlayer pairings. In this sense, 
only the intralayer pairings are important. We assume that the
two layers have the pairings $\Delta_i = g \sum_{{\bf k}} \langle c_{{\bf k}i\uparrow} c_{-{\bf k}i\downarrow} \rangle$. 
After the standard mean field treatment, we obtain
the following Bogoliubov-de Gennes (BdG) equation $\mathcal{H} = \sum_{{\bf k}} 
{1 \over 2} \Psi_{{\bf k}}^\dagger H_{8\times 8} \Psi_{{\bf k}}$, where
$H_{8\times 8} =  \begin{pmatrix}
                                \mathcal{H}_1 &  T \\
                                    T^\dagger & \mathcal{H}_2
                        \end{pmatrix}$, with $\mathcal{H}_1$ and $\mathcal{H}_2$ being 
			the Hamiltonian of the uncoupled layer, and $T = \text{diag}(-t, -t, t, t)$
                        is the bilayer coupling. $\mathcal{H}_i$ can be expressed as
\begin{equation}
        \mathcal{H}_i =
        \left(\begin{array}{cccc}
                \epsilon_{\bf k}^i -\Gamma & \rho_{\bf k}  & 0  & \Delta_i  \\ \smallskip
                                   \rho_{\bf k}^{\ast} & \epsilon_{\bf k}^i+\Gamma  & - \Delta_i &  0  \\ \smallskip
                                                        0 &  -\Delta_i^{\ast} & -\epsilon_{\bf k}^i+ \Gamma  & \rho_{\bf k}^{\ast} \\ \smallskip
                                       \Delta_i^{\ast} & 0  & \rho_{\bf k} &  -\epsilon_{\bf k}^i-\Gamma \\
                \end{array}\right),
\end{equation}
where $\rho_{{\bf k}} = -\alpha (k_y + ik_x)$. Here we have
chosen the Nambu basis $\Psi = (\Psi_{{\bf k}1}, \Psi_{{\bf
k}2})^T$, where $\Psi_{{\bf k}i} = (c_{{\bf k}i\uparrow}, c_{{\bf
k}i\downarrow}, c_{-{\bf k}i\uparrow}^\dagger, c_{-{\bf
k}i\downarrow}^\dagger)$. The corresponding thermodynamic
potential of this model can be written as
\begin{equation}
		        \Omega = {1\over 2} \sum_{{\bf k}}( \sum_{n < 0} 
				        E_{n{\bf k }} + 2\epsilon_{\bf k}^1 + 2\epsilon_{\bf k}^2) + {|\Delta_1|^2 + |\Delta_2|^2 \over g},
				\end{equation}
where in the first term sum over all occupied bands (eigenvalues $E_{n{\bf k}} < 0$) is assumed. Notice that the
divergence of the thermodynamic potential need to be regularized
using $g^{-1} = \sum_{{\bf k}} {1 \over {\bf k}^2/m + \varepsilon_\text{b}}$\cite{MG12, Randeria, Gang}, 
thus the binding energy $\varepsilon_\text{b}$ serves as the major parameter
to control the many-body interaction strength.

The minimization of the thermodynamic potential directly determines
all the roperties of the ground state, that is, we have
several equation sets --- total number equation $\partial \Omega/\partial
\mu = -n$ and order parameter equations $\partial \Omega / \partial
\Delta_i = 0$. In the following we numerically solve these equations
self-consistently. We define the Fermi momentum $k_\text{F} =
\sqrt{n\pi}$ and the Fermi energy $E_{\text{F}} = k_\text{F}^2/2m$, which 
are used to rescale the momentum and energy respectively during the numerical
simulation. Notice that $\Delta_1$ and $\Delta_2$ are complex numbers in principle,
however, in our model, these two parameters have the same global phase, thus
can be treated as real numbers simultaneously--- this point has been verified 
in all our numerical calculations. It can also be understood from two
basic facts. The single particle bands have inversion symmetry, that
is, the quantum state with momenta $\pm {\bf k}$ have the
same energy, thus we have uniform pairing. Moreover, the tunneling term 
can have minimal energy when these two order parameters have the same 
phase. This is also true in multilayer system when $t > 0$.

\begin{figure}
        \centering
        \includegraphics[width=2.8in]{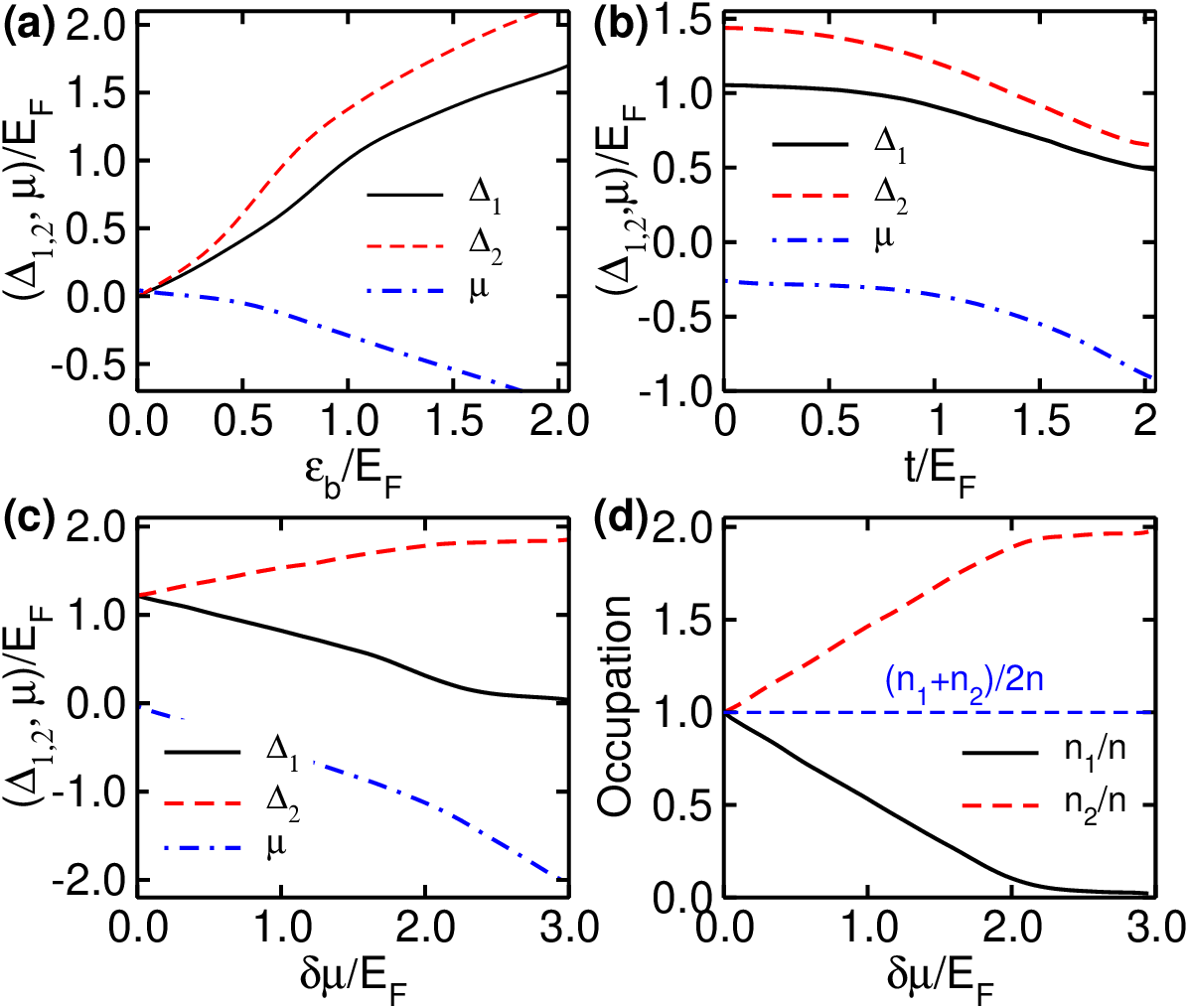}
        \caption{(Color online). The evolution of order parameters $\Delta_i$ and chemical potential $\mu$ as a function of binding energy (a), interlayer tunneling (b) and asymmetry (c). In (a) $t = 0.5E_\text{F}$, $\delta\mu = 0.5E_\text{F}$;  (b) $\varepsilon_\text{b} = 1.0E_\text{F}$, $\delta\mu = 0.5E_\text{F}$ and 
		(c) $t = 0.5E_\text{F}$, $\varepsilon_\text{b} = 1.0E_\text{F}$. The effect of asymmetry on the population occupation projected to each layer is 
                plotted in (d) with parameters from (c). In all subfigures $\Gamma = 1.0 E_\text{F}$.}
\label{fig-fig1}
\end{figure}

{\it BEC-BCS crossover}. We first show how binding energy, interlayer
tunneling and bilayer asymmetry affect the order parameters and
chemical potential. For other parameters --- SOC strength, Zeeman
field {\it etc}--- we have not observed qualitatively difference other than 
that reported in literatures\cite{MG11, MG112, MG113, MG12, JP, KS12, LH, MI, MI13, CQU13, WZ13, XJL13, CC13, RW12, 
Gang, HuHuia, HuHuib, XY, HLY, HLY2}. In Fig. \ref{fig-fig1} (a), we show the evolution of order 
parameters and chemical potential as a function of binding energy
$\varepsilon_\text{b}$. The chemical potential will change linearly with
respect to binding energy when $\varepsilon_\text{b}$ dominates in the Bose-Einstein condensation (BEC) 
regime. The difference between the two order parameters will be increased by the 
binding energy, which equivalently enhances the asymmetry effect. 
In Fig. \ref{fig-fig1} (b) we show an almost opposite effect to the order parameter from 
the interlayer tunneling, which tends to make the two layers have
the same occupation thus weaken the asymmetry effect.  In
Fig. \ref{fig-fig1} (c) we plot the effect of asymmetry on chemical potential and
order parameters while the corresponding occupations in each layer are plotted in Fig. \ref{fig-fig1} (d). 
As the increases of asymmetry, the cold atoms tend to populate in the layer with relative 
smaller chemical potential, thus we see that the order parameter and number of particle in this
layer will first increase linearly and finally saturate while in the other layer these values will 
gradually approach zero. Since we are focusing on the physics at zero temperature, all the
order parameters will never become zero --- this is not true at finite temperature\cite{MG11, MG112, MG113, MG12, JP, KS12, LH, MI, MI13, CQU13, WZ13, XJL13, CC13, RW12, Gang, HuHuia, HuHuib, XY, HLY, HLY2}. 
These features are essential to realize phase  transitions between different topological 
superfluids; see below.

\begin{figure}
        \centering
        \includegraphics[width=2.8in]{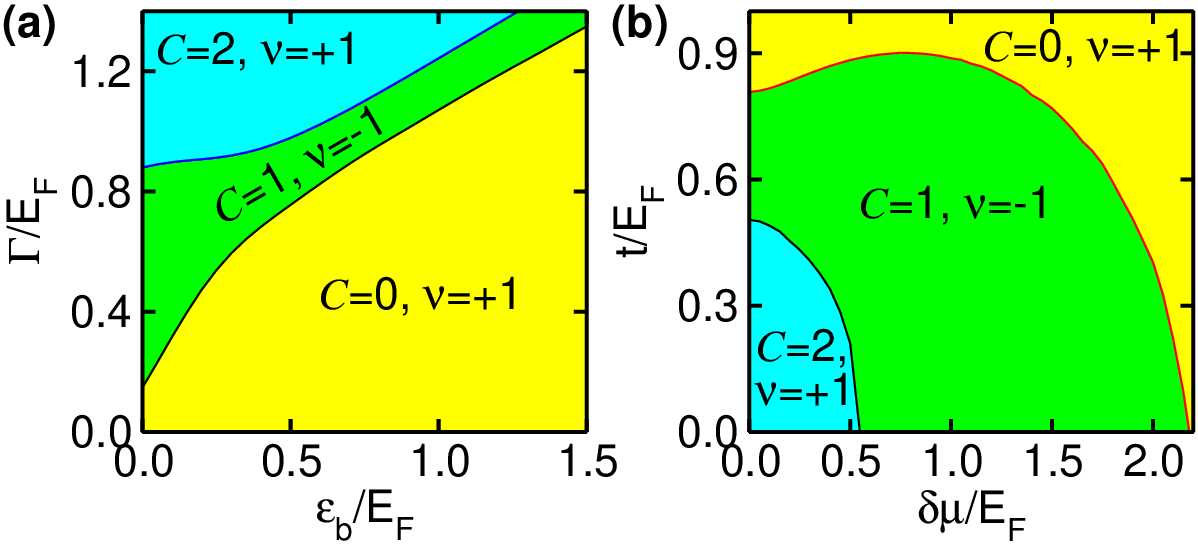}
        \caption{(Color online). Phase diagrams defined by Chern number $\mathcal{C}$ and sign of Pfaffian $\nu$. In (a), $t = \delta\mu = 0.5E_\text{F}$. In (b), $\varepsilon_\text{b} = 0.6E_\text{F}$, and $\Gamma = 0.97E_\text{F}$.}
\label{fig-fig2}
\end{figure}

{\it Phase diagram and topological phase transition}. The 
particle-hole symmetry is defined as $\Sigma = \Lambda K$
where $\Lambda = \rho_0 \otimes \tau_x \otimes \sigma_0$, with $\rho_0$, 
$\tau_x$ and $\sigma_0$ are the Pauli matrices act on bilayer space, particle-hole
space and spin space, respectively and $K$ is the complex conjugate
operator. We can easily verify that $\Sigma^2 = 1$, and 
$\Sigma H_{8\times 8} ({\bf k}) \Sigma^\dagger = - H_{8\times 8} (-{\bf
k})$. This is the only symmetry in our model thus the system belongs
to topological class $D$ with topological invariant $\mathbb{Z}$ in 2D\cite{ludwig, zhao}. 
We utilize three different methods to identify the topological
nature of the ground state. First, we can define a function
$W({\bf k}) = H_{8\times 8}({\bf k}) \Lambda $\cite{Ghosh, Pwave2}; following the 
basic definition of $\Sigma$, we have $W({\bf
k})^T = -W(-{\bf k})$. Thus $W(0)^T = -W(0)$ which is a real skew matrix. In this case, the sign of 
Pfaffian is defined as $\nu = \text{sign}(\text{Pf}(W(0)))$. Here $\nu$ is a topological protected
number because $\text{Pf}(W)= \pm \sqrt{\text{Det}(H(0))}$, which means 
that $\nu$ will never change sign upon deformation as long as the
energy gap is not closed. In previous proposals, $\nu = -1 (+1)$ 
corresponds to the topological nontrivial (trivial) phase, respectively\cite{Ghosh, Pwave2}. 
However, in our model, the sign of Pfaffian loses this meaning. In order to determine the 
topological invariant, we then proceed to calculate the Chern number of all the occupied bands
using $\mathcal{C} = \sum_{n < 0} \mathcal{C}_n$\cite{TKNN, Di}, where
$\mathcal{C}_n = {i \over 2\pi} \int dk_x dk_y \varepsilon^{ab} \langle \partial_{k_a} \psi_{n {\bf k}} | \partial_{k_b} \psi_{n {\bf k}} \rangle$. 
The calculated results using these two criteria are presented in Fig. \ref{fig-fig2}. These 
two definitions are consistent in the classification of topological phases in this model, that is, 
both topological invariants will undergo a discontinuous change across the critical boundaries. 

To understand these results, we can calculate the energy gap at zero momentum.
We find that the energy gap closing and reopening is determined by 
\begin{eqnarray}
\label{eq-boundary}
(\mu + \delta \mu)^2 + \mu^2 - 2\Gamma^2 + 2t^2 + \Delta_1^2 + \Delta_2^2 \pm  \sqrt{\mathcal{B}} = 0,
\end{eqnarray}
where $\mathcal{B} = (\Delta_1^2 - \Delta_2^2 - \delta \mu (2\mu + \delta \mu))^2 + 4t^2 ((\Delta_1 - \Delta_2)^2 + (2\mu + \delta \mu)^2)$. 
When $t = 0$ the above result reduces to $\Delta_i^2 +
\mu_i^2 = \Gamma^2$, which is a standard result in literatures\cite{MG11, MG12, solid1, solid2}. 
This result clearly indicates that the tunneling $t$ and asymmetry $\delta \mu$ directly enter 
the topological boundaries. We have verified that all the topological boundaries are
indeed determined by the above equation. Furthermore we have verified that all 
the bands are fully gapped when Eq. \ref{eq-boundary} is not fulfilled. In fact 
a finite Zeeman splitting is required to drive the topological phase transitions\cite{Comments1}.

\begin{figure}
        \centering
        \includegraphics[width=3.2in]{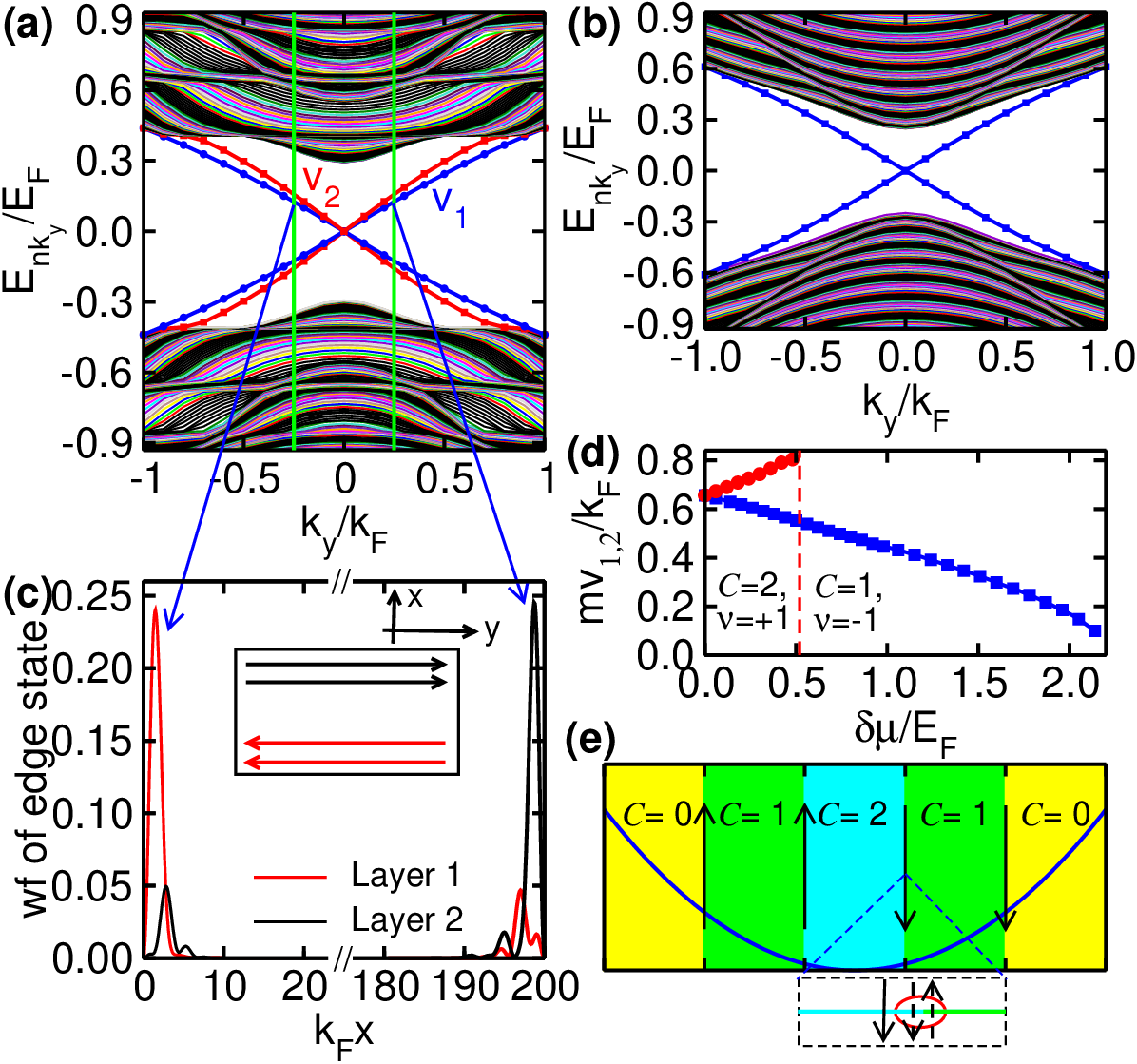}
        \caption{(Color online). Edge states in a strip geometry with Chern number $\mathcal{C} = 2$ (a) and $\mathcal{C} = 1$ (b). (c) is the total wavefunction of the edge states in each layer at momentum $k_y =\pm  0.25k_\text{F}$ (inset show the two states with the same momentum propagate at the same edge). (d) is the edge state velocities as a function of asymmetry. Parameters are $\Gamma=1.28E_\text{F}$ in (b) and $\Gamma=1.43E_\text{F}$ in (a) and (c), and the other parameters in (a) - (c) are: $\varepsilon_\text{b} = 1.2E_\text{F}$ and $t=\delta\mu=0.5E_\text{F}$. (d) show the velocities of the edge modes as a function of asymmetry under parameters $t = 0.2E_\text{F}$, $\varepsilon_\text{b} = 0.6E_\text{F}$ and $\Gamma = 0.97 E_\text{F}$. (e) Generalized bulk-edge correspondence in a trapped potential. The edge modes are spatially localized at the interface between two topological superfluids, and the number of edge modes $N = |\mathcal{C}_\text{L} - \mathcal{C}_\text{R}|$, where $\mathcal{C}_\text{L}$ and $\mathcal{C}_\text{R}$ are the Chern number of the superfluids at the left and right interface; all the other edge modes are gapped out by 
direct coupling due to their opposite chiralities.}
\label{fig-fig3}
\end{figure}

{\it Edge state and spin texture}. We then discuss the bulk-edge
correspondence in this model. We consider a strip geometry along $x$ direction with width $L = 200/k_\text{F}$ using
a hard wall boundary.  We expand the wavefunction using the plane waves, that is, 
$\psi(k_y) = \sum_{n < N_\text{c}} c_n e^{ik_y y} \sin(n\pi x/L)$, where $N_\text{c} = 200 $ is a basis cutoff\cite{Chan, Huhui}. The calculated results are presented
in Fig. \ref{fig-fig3}. We find that in all our calculations the number of edge 
modes exactly equals the Chern number $\mathcal{C}$, thus we have the bulk-edge correspondence. 
In Fig. \ref{fig-fig3} (a), we see that the two edge modes have linear dispersion at small $k_y$, but with different
velocities. These chiral edge modes can be described by $H = \sum_{i,k_y} v_i k_y \eta_{ik_y}^\dagger \eta_{ik_y}$\cite{XLQi}. 
The corresponding wavefunctions of the edge modes at $k_y = \pm 0.25k_\text{F}$ are presented in Fig. \ref{fig-fig3} (c), in which
we find that these two edge modes localized at the same edge propagate along the same direction (see inset of Fig. \ref{fig-fig3} (c)) 
--- this result is  in stark contrast to the solid system with time-reversal symmetry where the two edge modes localized at the same 
edge have opposite spin and counter propagate\cite{EdgeSH1, EdgeSH2, XLQi}. We also calculate the velocities of the 
edge modes as a function of asymmetry. When $\delta \mu < 0.52E_\text{F}$, one of the velocities increases almost linearly while 
the other mode decreases monotonically with respect to $\delta\mu$. When $2.14E_\text{F} > \delta \mu > 0.52E_\text{F}$, the 
edge mode with relative larger velocity disappears and only one edge mode exists since $\mathcal{C} = 1$. Strikingly, the evolution
of this velocity is a smooth function across the topological boundary; see Fig. \ref{fig-fig3} (d). In the trivial phase regime ($\delta \mu > 2.14E_\text{F}$),
no edge mode exists. In experiments for a typical $^{40}$K ($^6$Li) atoms with density $n =3.0 \times 10^{8}$/cm$^2$\cite{Li6, Li62, Li63}, 
we estimate $E_\text{F} = 1.2 $ (8.0) kHz, $k_\text{F}/m = 4.8$ (32.0) mm/s, thus both $v_{1}$ and $v_2$ should be around 3.0 (20.0) mm/s. 
These gapless modes are protected by a gap about 0.6$E_\text{F}$. These phases can be observed in the BEC-BCS 
crossover regime\cite{Alt, Bar, GM14} with $\ln k_\text{F} a_{\text{2D}} \sim 0 $, where $a_{\text{2D}}$ is the 2D scattering 
length. This is the third method to identify the bulk topology.

The stability of Majorana fermions (MFs) at $k_y=0$ is essential to understand the robustness of the edge modes. The MFs at the same edge
carry a definite chirality defined as $\Sigma^\dagger \psi_{nk_y=0} = \pm \psi_{nk_y=0}$. This chirality prohibits the 
direct fusion of the MFs. Consider a potential $\mathcal{V}_\text{d}$, which satisfies $\mathcal{V}_\text{d} = - \Sigma \mathcal{V}_\text{d} \Sigma^\dagger$. We can 
verify immediately that $\langle \psi_{i k_y=0} | \mathcal{V}_\text{d} | \psi_{j k_y=0}\rangle = \langle \psi_{i k_y=0}  | \Sigma \mathcal{V}_\text{d}\Sigma^\dagger | \psi_{j k_y=0}\rangle = - \langle \psi_{i k_y=0} |\mathcal{V}_\text{d} | \psi_{j k_y=0}\rangle = 0$, where 
$i \ne j$, thus direct coupling between any two MFs with the same chirality is forbidden. The edge states are robust in this sense. However, 
the condition is totally different in a trapped potential, see Fig. \ref{fig-fig3} (e), where edge states are spatially localized at the 
interface between two different topological superfluids. The direct fusion of MFs can occur due to their opposite chiralities, thus some of the edge modes will 
be gapped out by direct coupling. The remained number of edge modes equal their Chern number difference --- this picture is also verified in our
 calculation based on local density approximation. This is a generalized bulk-edge correspondence for large Chern number topological superfluids in a 
trapped potential. These edge modes, which can be accessed individually in experiments, provide important ground for measuring the local topology invariant of the superfluids.

\begin{figure}
        \centering
        \includegraphics[width=2.8in]{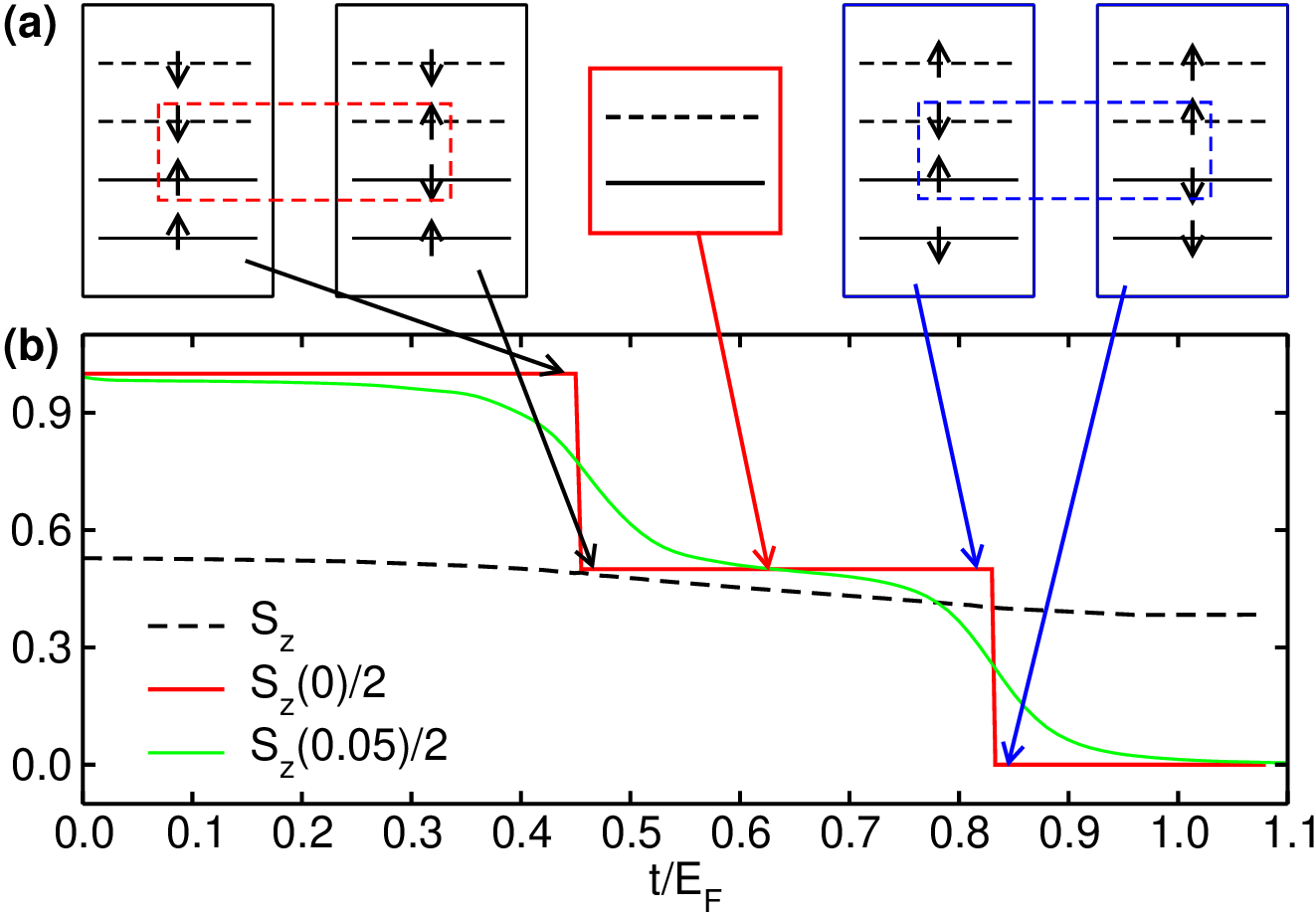}
		\caption{(Color online). (a) Band inversion and topological phase transitions. Across the critical points the spin polarization $S_\text{z}(0) = \langle c_{0\uparrow}^\dagger c_{0\uparrow} -c_{0\downarrow}^\dagger c_{0\downarrow} \rangle$ (marked by arrows) changes discretely due to interchange of symmetry between the occupied and unoccupied bands. The band inversions (see the third subfigure) among the occupied bands --- a typical feature in coupled layer system --- do not change the ground state topology, thus $S_\text{z}(0)$ is invariant. (b) Evolution of spin polarization at zero momentum $S_\text{z}(0)$ and total population imbalance $S_\text{z} = (n_{\uparrow} - n_{\downarrow})/n$ as a function of tunneling. $S_\text{z}(0.05)$ is the spin polarization at momentum 
				$|{\bf k}| = 0.05 k_\text{F}$.  Parameters are $\varepsilon_\text{b}=0.6E_\text{F}$, $\Gamma=0.97E_\text{F}$, $\delta\mu=0.2E_\text{F}$. }
\label{fig-fig4}
\end{figure}

The spin texture can be used to identify the topology of the ground state\cite{GM14, Foster13} due to band inversion; 
see Fig. \ref{fig-fig4} (a). In solid materials, the band inversion changes
the symmetry of conduction and valance bands\cite{Volk, Berg}, while in our model, the 
band inversion interchanges the spin polarization of the two touched bands. Across the 
critical point the energy gap first closes and then reopens, however, the spin texture 
will change discretely. A similar feature can be observed at small momentum, see the
curve $S_\text{z}(0.05)$ in Fig. \ref{fig-fig4}. This local spin texture at or near zero momentum therefore 
can be used to identify the topological phase transitions. To be more specific, we also calculate the total population 
imbalance across the critical point, which show that $S_\text{z}$ is a smooth function. This spin texture can 
be directly obtained from the time-of-flight imaging in experiments\cite{ZWIER, ALBA, SAUJ}.

{\it Large Chern number generation}. This idea can be straightforwardly generalized
to multilayer system, in which the topological invariant $\mathcal{C}$ equals any 
positive integer can be realized due to the additivity of the invariant in 
topological class D. The largest Chern number equals $N \mathcal{C}_0$, where 
$N$ is the number of coupled layers and $\mathcal{C}_0$ is the Chern number in a single uncoupled layer (here $\mathcal{C}_0 = 1$). 
Topological transitions between any two distinct topological phases can be realized by carefully engineering the system 
parameters. In this generation, several basic issues need to be remarked. (I) These basic observations will not be spoiled by the 
interlayer pairings, which are allowed in principle due to the weak wavefunction overlaps along the $z$ direction. 
As a proof-in-principle examination, we can include two arbitrary weak pairing
terms $\Delta_{\uparrow}' = g'\sum_{{\bf k}} \langle c_{{\bf k}1\uparrow}
c_{-{\bf k}2\downarrow}\rangle$ and $\Delta_{\downarrow}' = g' \sum_{{\bf k}}
\langle c_{{\bf k}1\downarrow} c_{-{\bf k}2\uparrow}\rangle$, where $g' \ll g$,
into $H_{8\times 8}$; we have verified that similar topological phases and
associated edge states can still be observed. (II) In the weak tunneling regime
the Chern number $\mathcal{C}$ can be treated as a sum of all Chern numbers
in each sublayer. However this simple picture is not true in the strong tunneling
regime. We illustrate this point using two counter examples. First, it is possible to
realize a topological superfluids by coupling two trivial layers in the strong 
tunneling regime. Furthermore, if the pairing in one of the layers is very weak, we can set 
the order parameter in this layer to zero without hurting the topological 
invariant of the whole system. This is because the proximity effect between a
normal layer and a  superfluid layer can greatly modify the topological invariant
of the whole system. Thus in the strong tunneling regime the Chern number $\mathcal{C}$ should be a 
feature of the whole coupled layer system. (III) During the topological phase transitions in multilayer 
system similar behavior of $S_\text{z}({\bf k})$ in Fig. \ref{fig-fig4} can still be observed due to band inversion.

To conclude we show using self-consistent calculation that
topological superfluids with large Chern number can be realized in coupled layer
system with spin-orbit coupling, Zeeman field and $s$-wave pairing.
This system admits the observation of topological phase transitions between
different topological superfluids, which are characterized by Chern number and sign of Pfaffian.
The edge states and spin texture across the topological boundaries 
also exhibit some intriguing features especially in a trapped potential. These large Chern 
number topological superfluids provide fertile grounds for exploring new quantum matters in context of ultracold atoms.

\textit{Acknowledgements}. B. H. is supported by Natural Science 
Foundation of Jiangsu Province under Grant No. BK20130424 and
National Natural Science Foundation of China under Grant No. 11275180.
M.G. and C. C are supported by Hong Kong RGC/GRF Projects (No. 401011 
and No. 2130352), University Research Grant (No. 4053072) and The 
Chinese University of Hong Kong (CUHK) Focused Investments Scheme.


\begin{thebibliography}{99}
\bibitem{SOC1} Y.-J. Lin, K. Jim\'{e}nez-Garc\'{\i}a, and I. B. Spielman, Nature (London) \textbf{471}, 83 (2011).
\bibitem{SOC2} R. A. Williams, M. C. Beeler, L. J. LeBlanc, K. Jim\'{e} nez-Garc\'{\i}a, and I. B. Spielman, Phys. Rev. Lett. \textbf{111}, 095301 (2013).
\bibitem{SOC3} J. Y. Zhang, S. C. Ji, Z. Chen, L. Zhang, Z. -D. Du, B. Yan, G. -S. Pan, B. Zhao, Y. -J. Deng, H. Zhai, S. Chen, and J. -W. Pan, Phys, Rev. Lett. \textbf{109}, 115301 (2012).
\bibitem{SOC4} L. W. Cheuk, A. T. Sommer, Z. Hadzibabic, T. Yefsah, W. S. Bakr, and M. W. Zwierlein, Phys. Rev. Lett. \textbf{109}, 095302 (2012).
\bibitem{SOC5} P. Wang, Z. -Q. Yu, Z. Fu, J. Miao, L. Huang, S. Chai, H.
Zhai, and J. Zhang, Phys. Rev. Lett. \textbf{109}, 095301 (2012).\qquad
\bibitem{SOC6} C. Qu, C. Hamner, M. Gong, C. Zhang, and P. Engels, Phys.
Rev. A \textbf{88}, 021604 (2013).
\bibitem{SOC7} Z. Fu, P. Wang, S. Chai, L. Huang, and J. Zhang, Phys. Rev. A \textbf{84}, 043609 (2011).
\bibitem{SOC8} C. Hamner, Yongping Zhang, M. A. Khamehchi, Matthew J. Davis, P. Engels, arXiv.1405.4048.

\bibitem{REV} V. Galitski and I. B. Spielman, Nature (London) \textbf{494}, 49 (2013).

\bibitem{FF64} P. Fulde and R. A. Ferrell, Phys. Rev. \textbf{135}, A550
(1964).
\bibitem{LO64} A. I. Larkin and Y. N. Ovchinnikov, Zh. Eksp. Teor. Fiz.
\textbf{47}, 1136 (1964).

\bibitem{MG11} M. Gong, S. Tewari, and C. Zhang, Phys. Rev. Lett. \textbf{107}, 195303 (2011).

\bibitem{MG112} Hui Hu, Lei Jiang, Xia-Ji Liu, and Han Pu, Phys. Rev. Lett. \textbf{107}, 195304 (2011).
\bibitem{MG113} Zeng-Qiang Yu and Hui Zhai, Phys. Rev. Lett. \textbf{107}, 195305 (2011).

\bibitem{MG12} M. Gong, G. Chen, S. Jia, and C. Zhang, Phys. Rev. Lett. \textbf{109}, 105302 (2012).
\bibitem{JP} J. P. Vyasanakere, S. Zhang, and V. B. Shenoy, Phys. Rev. B \textbf{84}, 014512 (2011).
\bibitem{KS12} K. Seo, L. Han, and C. A. R. S\'{a} de Melo, Phys. Rev. A
\textbf{85}, 033601 (2012).
\bibitem{LH} L. Han, and C. A. R. S\'{a} de Melo, Phys. Rev. A \textbf{85}, 011606(R) (2012).
\bibitem{MI} M. Iskin and A. L. Suba\c{s}i, Phys. Rev. Lett. \textbf{107},
063627 (2011).
\bibitem{MI13} M. Iskin and A. L. Suba\c{s}i, Phys. Rev. A \textbf{87},
050402 (2013).
\bibitem{CQU13} C. Qu, Z. Zheng, M. Gong, Y. Xu, L. Mao, X. Zou, G. Guo, and
C. Zhang, Nat. Commun. \textbf{4}, 2710 (2013).
\bibitem{WZ13} W. Zhang and W. Yi, Nat. Commun. \textbf{4}, 3710 (2013).
\bibitem{XJL13} X.-J. Liu and H. Hu, Phys. Rev. A \textbf{88}, 023622 (2013).
\bibitem{CC13} C. Chen, Phys. Rev. Lett. \textbf{111}, 235302 (2013).
\bibitem{RW12} R. Wei and E. J. Mueller, Phys. Rev. A \textbf{86}, 063604 (2012).
\bibitem{Gang}Gang Chen, Ming Gong, and Chuanwei Zhang, Phys. Rev. A \textbf{85}, 013601 (2012).
\bibitem{HuHuia} Hui Hu and Xia-Ji Liu,  New. J. Phys. \textbf{15}, 093037 (2013).
\bibitem{HuHuib} Xia-Ji Liu, Phys. Rev. A \textbf{88}, 043607 (2013).
\bibitem{XY} Y. Xu, C. Qu, M. Gong, and C. Zhang, Phys. Rev. A \textbf{89}, 013607 (2014).
\bibitem{HLY} Lianyi He and Xu-Guang Huang, Phys. Rev. Lett. \textbf{108}, 145302 (2012). 
\bibitem{HLY2} Lianyi He and Xu-Guang Huang, Phys. Rev. B \textbf{86}, 014511 (2012).

\bibitem{Pwave1} N. Read and D. Green, Phys. Rev. B \textbf{61}, 10267 (2000).
\bibitem{Pwave2} A. Yu. Kitaev, Physics-Uspekhi (supplement) \textbf{44}, 131 (2001).
\bibitem{Pwave3} G. Moore and N. Read,  Nuclear Physics B \textbf{360}, 362 (1991).
\bibitem{Pwave4} R. Matzdorf, Z. Fang, Ismail, Jiandi Zhang, T. Kimura, Y. Tokura, K. Terakura and E. W. Plummer, Science, \textbf{289}, 746 (2000).

\bibitem{CC1} A. M. Clogston, Phys. Rev. Lett. \textbf{9}, 266 (1962).
\bibitem{CC2} B. S. Chandrasekhar, Appl. Phys. Lett. \textbf{1}, 7 (1962).

\bibitem{ludwig} A. P. Schnyder, S. Ryu, A. Furusaki and A. W. W. Ludwig, Phys. Rev. B \textbf{78}, 195125 (2008). 
\bibitem{zhao} Y. X. Zhao and Z. D. Wang, Phys. Rev. Lett. \textbf{110}, 240404 (2013).

\bibitem{IQHE} R. Laughlin, Phys. Rev. B \textbf{23}, 5632 (1981).
\bibitem{Fang} Chen Fang, Matthew J. Gilbert, and B Andrei Bernevig, Phys. Rev. Lett. \textbf{112}, 046801 (2014).
\bibitem{Doru} Doru Sticlet and Frederic Piechon, Phys. Rev. B \textbf{87}, 115402 (2013).
\bibitem{FanZhang} Fan Zhang, Jeil Jung, Gregory A. Fiete, Qian Niu, and Allan H. MacDonald, Phys. Rev. Lett. \textbf{106}, 156801 (2011).
\bibitem{Volvik} G.E. Volovik, JETP \textbf{67}, 1804 (1988)


\bibitem{Randeria} M. Randeria, J.-M. Duan, and L.-Y. Shieh, Phys. Rev. Lett. \textbf{62}, 981 (1989).

\bibitem{Ghosh} P. Ghosh, J. D. Sau, S. Tewari, and S. Das Sarma, Phys. Rev. B \textbf{82}, 184525 (2010).

\bibitem{TKNN} D. J. Thouless, M. Kohmoto, M. P. Nightingale, and M. den Nijs, Phys. Rev. Lett. \textbf{49}, 405 (1982).
\bibitem{Di} D. Xiao, M.-C. Chang, and Q. Niu, Rev. Mod. Phys. \textbf{82}, 1959 (2010).

\bibitem{solid1} J. Alicea, Rep. Prog. Phys. \textbf{75}, 076501 (2012).
\bibitem{solid2} C. W. J. Beenakker, Annu. Rev. Con. Mat. Phys. \textbf{4}, 113 (2013).

\bibitem{Comments1} Analytical expression is hard to be obtained in the coupled layer system. However, in the limiting case that all layers have the same order parameter strength, $\Delta_i = \Delta$, we can prove exactly that the required minimal Zeeman splitting $|\Gamma|\ge |\Delta|$. This condition can be realized when $t^2 + \delta\mu^2 = \mu^2$. 

\bibitem{Chan} C. F. Chan and M. Gong, Phys. Rev. B \textbf{89}, 174501 (2014).
\bibitem{Huhui} H. Hu, L. Dong, Y. Cao, H. Pu, and X.-J. Liu, arXiv:1404.2442(2014).

\bibitem{XLQi} Xiao-Liang Qi and Shou-Cheng Zhang, Rev. Mod. Phys. \textbf{83}, 1057 (2011).

\bibitem{EdgeSH1} Y. Tanaka, A. Furusaki, and K. A.  Matveev, Phys. Rev. Lett. \textbf{106}, 236402 (2011).
\bibitem{EdgeSH2} C. Brune,	 A. Roth,	 H. Buhmann,	 E.  M. Hankiewicz,	 L. W. Molenkamp,	 J. Maciejko,	 X.-L. Qi	and S.-C. Zhang, Nature Physics \textbf{8}, 485 (2012).

\bibitem{Li6} K. M. O'Hara, S. L. Hemmer, M. E. Gehm, S. R. Granade, J. E. Thomas, Science, \textbf{298}, 2179 (2002).
\bibitem{Li62} Kirill Martiyanov, Vasiliy Makhalov, and Andrey Turlapov, Phys. Rev. Lett. \textbf{105}, 030404 (2010). 
\bibitem{Li63} Bernd Frohlich, Michael Feld, Enrico Vogt, Marco Koschorreck, Wilhelm Zwerger, and Michael Kohl, Phys. Rev. Lett. \textbf{106}, 105301 (2011).

\bibitem{Alt} A. Altmeyer, S. Riedl, C. Kohstall, M. J. Wright, R. Geursen, M. Bartenstein, C. Chin, J. Hecker Denschlag, and R. Grimm, Phys. Rev. Lett. \textbf{98}, 040401 (2007). 
\bibitem{Bar} M. Bartenstein, A. Altmeyer, S. Riedl, S. Jochim, C. Chin, J. Hecker Denschlag, and R. Grimm, Phys. Rev. Lett. \textbf{92}, 203201 (2004).

\bibitem{GM14} Y. Dong, L. Dong, M. Gong, and H. Pu, arXiv:1406.3821(2014).
\bibitem{Foster13} Matthew S. Foster, Maxim Dzero, Victor Gurarie, and Emil A. Yuzbashyan,  Phys. Rev. B \textbf{88}, 104511 (2013).

\bibitem{Volk} B. A. Volkov and O. A. Pankratov, JETP Lett, \textbf{42}, 145 (1985).
\bibitem{Berg} B. Andrei Bernevig, Taylor L. Hughes, Shou-Cheng Zhang, Science, \textbf{314}, 1757 (2006). 

\bibitem{ZWIER} L. W. Cheuk, A. T. Sommer, Z. Hadzibabic, T. Yefsah, W. S. Bakr, and M. W. Zwierlein, Phys. Rev. Lett. \textbf{111}, 095302 (2013).
\bibitem{ALBA} E. Alba, X. Fernandez-Gonzalvo, J. Mur-Petit, J. Pachos, and J. Garcia-Ripoll, Phys. Rev. Lett. \textbf{107}, 235301 (2011).
\bibitem{SAUJ} J. Sau, R. Sensarma, S. Powell, I. Spielman, and S. Das Sarma, Phys. Rev. B \textbf{83}, 140510(R) (2011).


\end{thebibliography}
\end{document}